\documentclass[preprint,aps,nofootinbib]{revtex4}
\usepackage{graphicx}
\usepackage{amsmath}
\usepackage{amsfonts}
\usepackage{amssymb}
\usepackage{color}%
\providecommand{\U}[1]{\protect\rule{.1in}{.1in}}

\newcommand{\f}{\begin{equation}}
\newcommand{\ff}{\end{equation}}
\newcommand{\fa}{\begin{eqnarray}}
\newcommand{\ffa}{\end{eqnarray}}

\begin{document}
\title{Thermal non-Gaussianity in Near-Milne universe}
\author{Jian-Pin Wu }
\email{jianpinwu@yahoo.com.cn}
\author{Yi Ling}
\email{yling@ncu.edu.cn} \affiliation{Center for Relativistic
Astrophysics and High Energy Physics, Department of Physics,
Nanchang University, 330031, China}

\begin{abstract}

The thermal non-Gaussianity in Near-Milne universe is investigated
in this letter. Through classifying thermal fluctuations into two
types, one characterized by a phase transition and the other
without phase transition, we show that for  fluctuations
undergoing a phase transition, the non-Gaussianity depends
linearly on $k$, but in non-phase transition case it is
proportional to $k^{-3\gamma},\gamma>0$. Moreover, if the phase
transition scenario is similar to that in holographic cosmology,
we find that the non-Gaussianity $f_{NL}^{equil}$ can reach ${\cal
O}(1)$ or larger than ${\cal O}(1)$ by fine tuning of the equation
of state $\omega_{C}$. Especially at the limit of
$\omega_{C}\rightarrow -5/3$, the non-Gaussianity can be very
large. On the other hand, for non-phase transition case, the
non-Gaussianity estimator $f_{NL}^{equil}$ is approximately
$f_{NL}^{equil}\simeq 5/108$ when the energy is sub-extensive,
especially when $\gamma \rightarrow 0^{+}$. However, when $\gamma
\rightarrow 4$, large non-Gaussianity can be obtained.

\end{abstract}
\maketitle

\section{Introduction}

The non-Gaussianity of the Cosmic Microwave Background (CMB) has
been playing a crucial role in the investigation of the very early
universe, since it has become a distinct feature among all kinds
of inflation models and their alternatives. For instance, for
single field slow roll inflation, the non-Gaussianity is too small
to be detected in near future\cite{inflationm,inflationo},
however, for models beyond single field slow roll inflation, the
non-Gaussianity large enough to be observable can be
 predicted\cite{DBI,K,Ghost,add,bilyth,bibartolo,bidvali,biling,bisenatore}. Recently,
WMAP 5-year data indicates that at 95\% confidence level, the
primordial non-Gaussianity parameters for the local and
equilateral models are in the region $-9<f_{\rm NL}^{local}<111$
and $-151<f_{\rm NL}^{equil}<253$, respectively \cite{WMAP5}. If
this result is confirmed by future experiments, such as the Planck
satellite, then it will be a great challenge to many inflation
models.

In this letter, we intend to explore the thermal non-Gaussianity in
Near-Milne universe. We find that thermal fluctuations in
Near-Milne universe can result in a large non-Gaussianity, similar
to the case in holographic cosmology \cite{biling}. Treating
thermal fluctuations as the origin of primordial fluctuations in
cosmology was advocated by Peebles \cite{Peebles}. Subsequently,
Magueijo {\it et al.} have more explicitly explored the
possibility that primordial thermal fluctuations, instead of
quantum fluctuations, might seed the structure of our universe
\cite{steph,rob,mag-pog}. Unfortunately, they find that the
spectral index of thermal fluctuations is either too red
($n_{s}=0$) or too blue ($n_{s}=4$), failing to generate a nearly
scale invariant spectrum \cite{thermalmilne, thermalloop,
thermalholography1}. Therefore, in order to produce nearly scale
invariant fluctuations, thermal scenarios call for new physics.
For instance, a thermal scenario with new physics effects to
change the equation of state of thermal matter can produce nearly
scale invariant spectrum. This happens in non-commutative
inflation \cite{steph,non-commutative1,non-commutative2}, and in
loop quantum cosmology \cite{ddrlqc,thermalloop}. It can also
occur in string gas cosmology with a Hagedorn phase\cite{Hagedorn}
or in holographic cosmology with a phase
transition\cite{thermalholography1}, in which the energy becomes
strongly non-extensive, specifically proportional to the area of
the system surface. Recently, it has also been studied that
thermal fluctuations in a non-singular bouncing cosmology may
produce scale invariant fluctuations as well as a large
non-Gaussianity \cite{bouncing}. Besides all above approaches,
postulating a mildly sub-extensive contribution to the energy
density in Near-Milne universe can lead to a scale-invariance
spectrum as well\cite{thermalmilne}. In this letter, we intend to
investigate the thermal non-Gaussianity in this context, following
the strategy developed in Ref. \cite{thermalChen}, which has also
been used to investigate the non-Gaussianity in string gas
cosmology \cite{thermalSGC} and holographic cosmology
\cite{biling}.

The outline of this letter is following. In section \textrm{II} we
firstly calculate the thermal non-Gaussianity in Near-Milne
universe by classifying the thermal fluctuation into phase
transition type and non-phase transition type.  We summarize our
results and discuss some open questions in section \textrm{III}.

\section{Thermal non-Gaussianity in Near-Milne universe}

In cosmology the primordial scale invariance due to thermal
fluctuation is usually implemented through two ways, phase
transition or non-phase transition. For instance, in holographic
cosmology as well as string gas cosmology, the thermal fluctuation
undergoes a phase transition, in which the final spectrum is
proportional to the equal temperature spectrum. On the other hand,
in semi-classical loop quantum cosmology there is non-phase
transition, in which comoving scales $k$ start thermalized inside
the horizon, and then there exists a mechanism pushing sub-horizon
thermal modes outside the horizon, where they freeze and become
non-thermal. While in Near-Milne universe, the primordial scale
invariance can be realized through either phase transition or
non-phase transition \cite{thermalmilne}. Therefore, in this letter,
we will consider the thermal non-Gaussianity in Near-Milne universe
by classifying thermal fluctuations into phase transition type and
non-phase transition type.

Before discussing the non-Gaussianity, a key issue which has to be
addressed is on what scale the initial conditions should be imposed.
In Ref.\cite{thermalChen}, the thermal horizon $R$ is considered as
a free parameter. Then, if thermal horizon is smaller than Hubble
horizon scale at the horizon crossing, a large non-Gaussianity can
be obtained. In Ref.\cite{non-commutative2}, the thermal correlation
length $R=T^{-1}$ is adopted, which is a lower bound and so will
lead to larger non-Gaussianity. In our work, we will adopt the
Hubble scale $R=H^{-1}$, beyond which causality prohibits local
causal interactions \cite{causality}. Therefore we will calculate
the spectrum at $R=H^{-1}$, i.e. $k=a/R=aH$.

\subsection{the case of phase transition}

In this subsection, we will firstly calculate the 2-point
correlations, 3-point correlations and the power spectrum, then
give the thermal non-Gaussianity in Near-Milne universe.

Fluctuations in a thermal ensemble can be derived from the
thermodynamic partition function
\begin{equation}
  Z=\sum_r e^{-\beta E_r}~,
\end{equation}
where the summation runs over all states. $E_{r}$ is the energy of
state $r$, and $\beta=T^{-1}$.

Now, to derive a scale invariant spectrum we adopt the assumption
proposed in \cite{thermalmilne} that both energy and entropy are
sub-extensive such that they can be expressed as
\begin{equation}\label{CasimirU}
U = \rho_{C}(T)V^{1-\gamma},
\end{equation}
\begin{equation}\label{CasimirS}
S = s_{C}(T)V^{1-\gamma},
\end{equation}
with $\gamma>0$. Then, the 2-point correlation function for the
energy density fluctuations is given by
\begin{equation}\label{rhomilne2}
  \langle\delta\rho^2\rangle=\frac{\langle\delta
  U^2\rangle}{V^{2}}=\frac{1}{V^2}\frac{d^2 \log Z}{d\beta^2}=-\frac{1}{V^{2}}\frac{d\langle
  U\rangle}{d\beta}=\frac{T^{2}\rho'_{c}}{V^{1+\gamma}}~.
\end{equation}

Similarly, the 3-point correlation function can be expressed as
\begin{equation}\label{rhomilne3}
  \langle\delta\rho^3\rangle=\frac{\langle\delta
  U^3\rangle}{V^{3}}=-\frac{1}{V^3}\frac{d^3\log
  Z}{d\beta^3}=\frac{1}{V^3}\frac{d^2\langle U\rangle}{d\beta^2}=\frac{T^{3}(2\rho'_{c}+T\rho''_{c})}{V^{2+\gamma}}~.
\end{equation}

Performing the Fourier transformation, the density fluctuations
$\delta\rho_{k}$ in momentum space can be related to the
fluctuation $\delta\rho$ in position space by
\begin{equation}\label{rhok}
  \delta\rho_{k}=k^{-{3\over2}}\delta\rho~.
\end{equation}

Now to obtain the power spectrum of fluctuations we consider the
perturbation of $FRW$ metric. In longitudinal gauge (see \cite
{longitudinal1,longitudinal2,longitudinal3}), and in the absence
of anisotropic matter stress, the metric takes the form
\begin{equation}\label{metric}
  ds^{2}=a^{2}(\eta)[-d\eta^{2}(1-2\Phi)+(1+2\Phi)dx^{2}]~,
\end{equation}
where $\Phi$ represents the fluctuations in the metric. We assume
that the perturbations are deep in the horizon, which means $k\gg
H$. Then the perturbation equation of the metric in
Eq.(\ref{metric}) may be reduced to the Poisson equation
\begin{equation}
  k^{2}\Phi_{k}=4 \pi G a^{2}\delta\rho_{k}~.
\end{equation}

Therefore, the 2-point correlation function for $\Phi_{k}$ at
$k=a/R$ is given by
\begin{equation}\label{Phimilne2}
  \langle\Phi_{k}^2\rangle=(4\pi
  G)^{2}R^{1-3\gamma}T^{2}\rho'_{C}k^{-3}~,
\end{equation}

while the 3-point correlation function is
\begin{equation}\label{Phimilne3}
  \langle\Phi_{k}^3\rangle=(4\pi G)^{3}R^{-3\gamma}T^{3}(2\rho'_{C}+T\rho''_{C})k^{-\frac{9}{2}}~.
\end{equation}

At first, consider the  case with a fixed temperature, the
equal-time power spectrum is
\begin{equation}\label{powermilne1}
  P_{\Phi}=8G^{2}T^{2}\rho'_{C}R^{1-3\gamma}=8G^{2}T^{2}\rho'_{C}(\frac{a}{k})^{1-3\gamma}~.
\end{equation}

We can immediately see that for $\gamma=1/3$ the fixed temperature
power spectrum is scale-invariant. We notice that in this case,
the energy scales like $R^{2}$, which is similar to the case in
thermal holographic cosmology and string gas cosmology. As a
matter of fact, it is this modification that make it possible to
obtain scale invariant spectrum in all scenarios above.

Next, we calculate the thermal non-Gaussianity in Near-Milne
universe. Note that $\Phi$ perturbs CMB through the so-called
Sachs-Wolfe effect \cite{Sachs-Wolfe}. However, it is useful to
introduce a second variable, $\zeta$, which is the primordial
curvature perturbation on comoving hypersurfaces
\cite{Bardeen80,Bardeen83}. Then the non-Gaussianity estimator
$f_{NL}^{equil}$ can be calculated theoretically by
\begin{equation}\label{fdefine}
  f_{NL}^{equil} = \frac{5}{18}k^{-\frac{3}{2}}\frac{\langle {\zeta}_{k}^{3}\rangle}{\langle{\zeta}_{k}^{2}\rangle \langle{\zeta}_{k}^{2}\rangle}~.
\end{equation}

The variables $\Phi$ and $\zeta$ are related by
\begin{equation}\label{Phizeta1}
  \zeta=\Phi-\frac{H}{\dot{H}}(\dot{\Phi}+H\Phi)~.
\end{equation}

In general the variable $\zeta$ remains nearly constant at
super-horizon scales for adiabatic fluctuations but $\Phi$ not
\cite{longitudinal1}. However, if the equation of state is
constant, then $\Phi$ also remains constant at super-horizon.
Therefore the relation in (\ref{Phizeta1}) reduces to
\cite{Kodama},
\begin{equation}\label{Phizeta2}
  \zeta=\frac{5+3\omega}{3+3\omega}\Phi.
\end{equation}

We point out that the primordial curvature variable $\zeta$ is
independent of $\omega$, but the variable $\Phi$, which perturbs
the CMB, varies with the change of the equation of state $\omega$. For more detailed
discussion on the variables $\zeta$ and $\Phi$, we refer to
\cite{longitudinal1,NGReview,non-commutative2}.

Finally, combining Eqs. (\ref{Phimilne2}), (\ref{Phimilne3}),
(\ref{fdefine}) and (\ref{Phizeta2}), the non-Gaussianity
estimator $f_{\rm NL}^{equil}$ can be expressed as
\begin{equation}\label{fmilne1}
  f_{\rm NL}^{equil} =\frac{5}{6}\frac{1+\omega_{C}}{5+3\omega_{C}}\frac{(2\rho'_{C}+T\rho''_{C})}{4\pi GR^{2-3\gamma}T(\rho'_{C})^{2}}=\frac{5}{6}\frac{1+\omega_{C}}{5+3\omega_{C}}\frac{(2\rho'_{C}+T\rho''_{C})k}{4\pi GT(\rho'_{C})^{2}a}~.
\end{equation}

The condition for modes exiting and re-entering the Hubble radius
is $k=aH=a_{r}H_{r}$ and $k_{0}=a_{0}H_{0}$, where $H$, $a$ and
$H_{r}$, $a_{r}$ represent the values of exiting or re-entering
the Hubble radius, respectively, whereas $H_{0}$, $T_{0}$, and $k_{0}$ is today's values. For simplicity, we only consider all the process occurs in radiation-dominated era such that the
relation between the scale factor and the temperature is
$\frac{a_{c}}{a_{0}}=\frac{T_{0}}{T_{c}}$. As a result, the
non-Gaussianity estimator $f_{\rm NL}^{equil}$ becomes
\begin{equation}\label{f2}
  f_{\rm NL}^{equil} = \frac{5}{6}\frac{1+\omega_{C}}{5+3\omega_{C}}\frac{(2\rho_{C}'+T\rho_{C}'')}{4 \pi G(\rho_{C}')^{2}}\frac{H_{0}}{T_{0}}\frac{k}{k_{0}}=\frac{5}{6}\frac{1+\omega_{C}}{5+3\omega_{C}}\frac{(2\rho_{C}'+T\rho_{C}'')}{4 \pi G(\rho_{C}')^{2}}\times 10 ^{-30}\frac{k}{k_{0}}~,
\end{equation}
where $a_{c}$ is the scale factor during the phase transition and we have
neglected the variation in $a_{c}$.

From above equation, we find that if there exist a phase transition
process to render the above process, as described in holographic
cosmology, then all the things will be almost the same. For
instance, if $\omega_{C}\rightarrow -1$, the non-Gaussianity will be
suppressed as in usual inflationary scenario. If the matter is
phantom-like, the non-Gaussianity $f_{NL}^{equil}$ can reach ${\cal
O}(1)$ or larger than ${\cal O}(1)$ by fine tuning of the equation
of state $\omega_{C}$. Especially at the limit
$\omega_{C}\rightarrow -5/3$, the non-Gaussianity can be very large.
Moreover, the non-Gaussianity estimator $f_{NL}^{equil}$ depends
linearly on the mode $k$.

\subsection{the case of non-phase transition}

In this subsection we consider thermal fluctuations without
undergoing a phase transition process. Firstly we address the
issue on thermodynamical constraints. It has been noticed
\cite{constraint1,constraint2,constraint3} that the state
parameter $\omega$ and the Stephan-Boltmann law are linked by a
thermodynamic relation (we refer to Refs.
\cite{thermalloop,thermalmilne,thermalChen}). Suppose that both
energy and entropy are extensive, if the state parameter $\omega$
is a constant and $\rho=AT^{m}$, then one can obtain a
thermodynamic constraint
\begin{equation}\label{arhomilne}
  m=1+\frac{1}{\omega}.
\end{equation}

It has been realized in \cite{thermalloop,thermalmilne} that it is
this constraint that prevents us from achieving a scale invariant
($n_{s}=1$) spectrum for the energy density perturbation, but a
spectrum with $n_{s}=4$
 regardless the value of $\omega$. However,
in Near-Milne universe with sub-extensive energy and pressure, for
$p_{C}=\omega_{C}\rho_{C}$ and $\rho_{C}\propto T^{m_{C}}$, the
thermodynamic constraint can be modified as \cite{thermalmilne},
\begin{equation}\label{constraintCasimir}
m_{C} = 1+\frac{1-\gamma}{\omega_{C}}.
\end{equation}

Next, with such a modified constraint we derive the condition for
scale-invariance and calculate the thermal non-Gaussianity. We
assume that the modes freeze for $k=aH$, then using Eq.
(\ref{powermilne1}) and Friedmann equation $H\propto
\sqrt{\rho_{C}}$, outside the horizon we have
\begin{equation}\label{PTmilne1}
  \frac{d\ln P_{\Phi}}{d\ln
  T}=2+T\frac{\rho''_{C}}{\rho'_{C}}-\frac{1-3\gamma}{2}T\frac{\rho'_{C}}{\rho_{C}}.
\end{equation}

Using the relation $\rho_{C}\propto T^{m_{C}}$, the above equation
can be re-expressed as
\begin{equation}\label{PTmilne2}
  \frac{d\ln P_{\Phi}}{d\ln T}=1+\frac{m_{C}}{2}(1+3\gamma).
\end{equation}

Furthermore, since $k=aH\propto a\sqrt{\rho_{C}}$ and $a\propto
\rho_{C}^{-\frac{1}{3(1+\omega_{C})}}$, it can be found that
\begin{equation}\label{PTmilne3}
  \frac{d\ln k}{d\ln
  T}=\frac{3\omega_{C}+1}{6(1+\omega_{C})}\frac{T\rho'_{C}}{\rho_{C}}=\frac{(3\omega_{C}+1)m_{C}}{6(1+\omega_{C})}.
\end{equation}

Combining Eq.(\ref{PTmilne2}) and Eq.(\ref{PTmilne3}), we can
derive the spectral index as
\begin{equation}\label{indexmilne}
  n_{s}-1=\frac{d\ln P_{\Phi}}{d\ln
  k}=3\frac{(1+\omega_{C})[2+m_{C}(1+3\gamma)]}{m_{C}(3\omega_{C}+1)}.
\end{equation}
Therefore the condition for scale-invariance is
\begin{equation}\label{conditionmilne}
  m_{C}=-\frac{2}{1+3\gamma}.
\end{equation}

With the use of the modified constraint (\ref{constraintCasimir}),
the condition for scale-invariance can be re-expressed as
\begin{equation}\label{conditionconstraintCasimir}
  \omega_{C}=-\frac{(1-\gamma)(1+3\gamma)}{3(1+\gamma)}\approx -\frac{1}{3}(1+\gamma).
\end{equation}
In hence for sub-extensive energy with $\gamma>0$ we have
$\omega_{C}<-1/3$, which can provide an acceleration mechanism.

Now, we consider the non-Gaussianity. Similarly, the
non-Gaussianity estimator $f_{NL}^{equil}$ takes the form
\begin{equation}\label{fmilne2}
  f_{NL}^{equil}=\frac{5}{6}\frac{1+\omega_{C}}{5+3\omega_{C}}\frac{(2\rho'_{C}+T\rho''_{C})}{4\pi GR^{2-3\gamma}T(\rho'_{C})^{2}}~.
\end{equation}

As discussed previously if we take $R=H^{-1}$ and employ Friedmann
equation, then the non-Gaussianity estimator $f_{NL}^{equil}$ can
be expressed as
\begin{equation}\label{fmilne3}
  f_{NL}^{equil} =\frac{5(1+\omega_{C})}{9(5+3\omega_{C})}\frac{\rho_{C}(2\rho'_{C}+T\rho''_{C})}{T(\rho'_{C})^{2}}H^{-3\gamma}~.
\end{equation}

Using Eq.(\ref{conditionmilne}),(\ref{conditionconstraintCasimir})
and $\rho_{C}=AT^{m_{C}}$, we can further simplify
Eq.(\ref{fmilne3}) as
\begin{equation}\label{fmilne4}
  f_{NL}^{equil} =\frac{5(2-\gamma)(1-3\gamma)}{54(4-\gamma)}H^{-3\gamma}~.
\end{equation}

Similarly, the above equation can be approximately re-expressed as
\begin{equation}\label{fmilne5}
  f_{NL}^{equil} =\frac{5(2-\gamma)(1-3\gamma)}{54(4-\gamma)}\left(\frac{H_{0}}{T_{0}}T\right)^{-3\gamma}\left(\frac{k}{k_{0}}\right)^{-3\gamma}=\frac{5(2-\gamma)(1-3\gamma)}{54(4-\gamma)}\times(10^{-30}T)^{-3\gamma}\left(\frac{k}{k_{0}}\right)^{-3\gamma}~.
\end{equation}

From the above equation, we can immediately find that the
non-Gaussianity is proportional to $k^{-3\gamma}$ with $\gamma>0$.
It is very different from the cases in string gas cosmology,
holographic cosmology or phase transition case of Near-Milne
universe, in which the non-Gaussianity depends linearly on $k$.
However, in the non-phase transition case, the dependence on $k$ is
correlated with the value of parameter $\gamma$.

When the energy is sub-extension, especially when $\gamma
\rightarrow 0^{+}$, the non-Gaussianity estimator $f_{NL}^{equil}$
is approximately evaluated as $f_{NL}^{equil}\simeq 5/108$. However,
if we relax this condition and consider that the energy is not
sub-extensive, we find that as $\gamma \rightarrow 1/3$ and $\gamma
\rightarrow 2$, the non-Gaussianity is suppressed, whereas as
$\gamma \rightarrow 4^{-}$, the large and positive non-Gaussianity
can be obtained. Therefore, through an appropriate choice of the
parameter $\gamma$, we can make the non-Gaussianity
$f_{NL}^{equil}\sim {\cal O}(1)$ or larger than ${\cal O}(1)$. In
addition, we must notice that if $\gamma >4$, the non-Gaussianity
will be negative.

\section{Conclusion and discussion}

In this letter we have discussed the non-Gaussianity in Near-Milne
universe. We find that in the phase transition case, the
non-Gaussianity depends linearly on $k$ at fixed temperature, but in
non-phase transition case it is proportional to
$k^{-3\gamma},\gamma>0$. Furthermore, if the phase transition
scenario is analogous to the one described in holographic cosmology,
then all the results are almost the same. More discussions can be
found in Ref.\cite{biling}. However, we must point out that the
mechanism of phase transition plays an important role in this
scenario. If the mechanism is different from the case in holographic
cosmology, all the things could change. Currently how to implement
the phase transition is still an open question. We expect that we
can make further investigations on this issue in the future.
Moreover, it is worthwhile to notice that the non-Gaussianity
estimator $f_{NL}^{equil}$ also depends on the formulation of the
Stephan-Boltzmann law in this case.

In the case of non-phase transition, we have demonstrated that
 by appropriate choice of the parameter $\gamma$, we can
make the non-Gaussianity $f_{NL}^{equil}\sim {\cal O}(1)$ or larger
than ${\cal O}(1)$. If the energy is sub-extensive, especially when
$\gamma \rightarrow 0^{+}$, the non-Gaussianity estimator is
approximately $f_{NL}^{equil}\simeq 5/108$. However, when $\gamma
\rightarrow 4^{-}$, the large non-Gaussianity can be obtained.
Furthermore, if $\gamma>4$, the non-Gaussianity will be negative. In
addition, we note that when $\gamma\rightarrow 4$, the effective
state parameter $\omega_{C}\rightarrow -5/3$. As pointed out in Ref.
\cite{biling}, when $\omega\rightarrow -5/3$, $\zeta\rightarrow 0$,
so the non-Gaussianity estimator $f_{NL}^{equil}\rightarrow \infty$.
Here we can see this from the expression (\ref{fdefine}) of the
non-Gaussianity estimator $f_{NL}^{equil}$ and the relation
(\ref{Phizeta2}) as well.

Finally, we would like to point out that in order to bypass the
above process, we have to accept negative temperatures
\cite{constraint3,NT}. For more detailed discussion, we refer to
Ref.\cite{thermalmilne}. In the future, we will consider the
thermal non-Gaussianity in semi-classical loop cosmology
\cite{loopjianpinwu}.

\section*{Acknowledgement}

This work is partly supported by NSFC(Nos.10663001,10875057),
JiangXi SF(Nos. 0612036, 0612038), Fok Ying Tung Education
Foundation(No. 111008) and the key project of Chinese Ministry of
Education(No.208072). We also acknowledge the support by the
Program for Innovative Research Team of Nanchang University.

\end{document}